\documentclass[12pt]{article}

\begin{document}

\title{\bf \Large Solutions by radicals at singular values $k_{N}$\\
from new class invariants for $N\equiv3~{\rm mod}~8$}

\author{David Broadhurst\thanks{Department of Physics and Astronomy,
The Open University, Milton Keynes, MK7 6AA, United Kingdom,
{\tt D.Broadhurst@open.ac.uk}.}}

\date{\today}

\maketitle

\abstract{For square-free $N\equiv3$~mod~8 and $N$ coprime
to 3, I show how to reduce the singular value $k_N$ to
radicals, using a novel pair $[f,g]$ of real numbers that
are algebraic integers of the Hilbert class field of
$Q(\sqrt{-N})$. One is a class invariant of modular
level 48, with a growth $g=\alpha(N)\exp(\pi\sqrt{N}/48)+o(1)$,
where $\alpha(N)\in[-\sqrt2,\sqrt2]$ is uniquely determined by the residue
of $N$ modulo 64. Hence $g$ is a very economical generator of
the class field. For prime $N\equiv3$~mod~4, I conjecture
that the Chowla--Selberg formula provides an algebraic {\em unit}
of the class field and determine its minimal polynomial
for the 155 cases with $N<2000$. For $N=2317723$, with
class number $h(-N)=105$, I compute the minimal polynomial of
$g$ in 90 milliseconds. Its height is smaller
than the {\em cube} root of the height of the generating
polynomial found by the double eta-quotient method of {\em Pari-GP}.
I reduce the complete elliptic integral $K_{2317723}$ to radicals
and values of the $\Gamma$ function, by determining the
Chowla--Selberg unit and solving the septic, quintic and
cubic equations that generate sub-fields of the class field.
I conclude that the residue 3 modulo 8, initially discarded
in elliptic curve primality proving, outperforms the residue 7.}

\section{Introduction}

The $N$th singular value~\cite{Watson-sing,B-C-Z}
is the algebraic number $k_N\in[0,1]$ for which
\begin{equation}
{\rm AGM}\left(1,\sqrt{1-k_N^2}\right)=\sqrt{N}\,{\rm AGM}(1,k_N)
\label{k-N}
\end{equation}
where the arithmetic-geometric mean (AGM) is obtained by
iterating the rapidly convergent process~\cite{AGM}
${\rm AGM}(a,b)={\rm AGM}\left((a+b)/2,\sqrt{a b}\right)$.
For square-free $N\equiv3$~mod~8, with $N$ coprime
to 3,
\begin{equation}
k_N^2=\frac12-\sqrt{\frac14-\frac{16}{r^{24}}}
\label{r}
\end{equation}
is determined by an algebraic number $r>2^{\frac14}$
that is given by a Weber function~\cite{Weber,Atkin-M,Y-Z,Hajir-V,LMS,Schertz}
and has a minimal polynomial of degree $3h$, where
$h=h(-N)$ is the class number of the imaginary quadratic field $Q(\sqrt{-N})$.

For square-free $N\equiv3$~mod~4,
the complete elliptic integral
\begin{equation}
K_N=\int_0^1\frac{{\rm d}x}{\sqrt{(1-x^2)(1-k_N^2x^2)}}
=\frac{\pi}{2}\,\frac{1}{{\rm AGM}\left(1,\sqrt{1-k_N^2}\right)}
\label{K-N}
\end{equation}
is reducible to the $\Gamma$ values~\cite{C-S-1,Chowla-S,Zucker} in
\begin{equation}
G_N=\prod_{k=1}^{N}\left[\Gamma\left(\frac{k}{N}\right)\right]
^{\left(\frac{-N}{k}\right)}
\label{G-N}
\end{equation}
with exponents given by the Legendre--Jacobi--Kronecker symbol
$\left(\frac{-N}{k}\right)$.  For $N>3$,
this reduction takes the form
\begin{equation}
K_N=\left(\frac{r}{2}\right)^2\sqrt{\frac{2\pi}{N}
\left(\lambda^4G_N\right)^{\frac{1}{h}}}
\label{w}
\end{equation}
where $\lambda>0$ is an algebraic number.
As noted in~\cite[Eq.~8]{C-S-1}, $\lambda=1$ when $h(-N)=1$.
Moreover, I conjecture in this paper
that $\lambda$ is an algebraic {\em unit} of the Hilbert class field
when $h(-N)$ is odd, i.e.~for prime $N>3$ congruent to 3 modulo 4.

I shall describe how $r$ and $\lambda$ were reduced to radicals
in the case $N=2317723$, with class number $h(-N)=105$.
To achieve this reduction,
I construct, in Section~4.2, a pair of class invariants,
one of which appears to outperform the {\tt quadhilbert}
procedure of {\em Pari-GP}, in regard of the economy
with which it generates the class field.

\section{Chowla--Selberg formula}

It is not necessary to compute $N$ values of the
$\Gamma$ function to evaluate $G_N$ at high precision. Instead
we may use $h$ values of the Dedekind eta function
\begin{equation}
\eta(z)=\exp(\pi{\rm i}z/12)\prod_{k=1}^\infty(1-\exp(2\pi{\rm i}k z))
=\sum_{n=-\infty}^\infty(-1)^n\exp((6n+1)^2\pi{\rm i}z/12)
\label{eta}
\end{equation}
to evaluate $G_N$ using
the Chowla--Selberg formula~\cite[Eq.~2, p.~110]{Chowla-S}
\begin{equation}
\prod_{k=1}^{N}\left[\Gamma\left(\frac{k}{N}\right)\right]
^{\left(\frac{-N}{k}\right)}
=(2\pi N)^h\prod_{[a,b,c]\in H}\frac{1}{a}
\left|\eta\left(\frac{b+\sqrt{-N}}{2a}\right)\right|^4
\label{C-S}
\end{equation}
for square-free $N\equiv3$~mod~4 and $N>3$.
For other cases, including non-fundamental discriminants,
see~\cite{Huard-K-W}.
In~(\ref{C-S}), the product runs over the strict equivalence classes
$[a,b,c]$ of primitive integral binary quadratic forms
$a x^2+b x y+c y^2$ with discriminant $b^2-4a c=-N$.
These equivalence classes form an Abelian
group $H$, by Gauss's composition of quadratic forms, and the order of
$H$ is the class number $h=h(-N)$. It is remarked in~\cite{Huard-W}
that publication of this striking formula was delayed for 18 years,
between its discovery at the time of the Chowla--Selberg
paper~\cite{C-S-1} of 1949 and its appearance in the
Selberg--Chowla paper~\cite{Chowla-S} of 1967. For precursors
of this formula, see~\cite{J-N}.

\subsection{A conjecture for prime discriminants}

For square-free positive $N\equiv3$~mod~4, I define
\begin{equation}
\lambda=\prod_{[a,b,c]\in H}a^{\frac14}\left|\frac
{\eta\left(\frac{1+\sqrt{-N}}{2 }\right)}
{\eta\left(\frac{b+\sqrt{-N}}{2a}\right)}\right|
\label{lambda}
\end{equation}
where the product runs over the equivalence classes
for discriminant $b^2-4a c=-N$.\\[5pt]
{\bf Conjecture 1}: For prime $N\equiv3$~mod~4, $\lambda$ is a unit
of the Hilbert class field of $Q(\sqrt{-N})$.\\[5pt]
Remarks:\begin{enumerate}
\item I have verified this in the 155 cases with $N<2000$.
\item For each of these cases, the minimal polynomial $L(x)$ of
$\lambda$ is available\footnote{From the directory
{\tt http://paftp.open.ac.uk/pub/staff\_ftp/dbroadhu/K2317723/}~.}
in a file {\tt lambdaCS.txt} which is read by {\tt lambdaCS.gp}
with output in {\tt lambdaCS.out} that confirms,
at a precision of 15,000 digits, that $L(\lambda)=0$
and that $L$ generates the same field as the {\tt quadhilbert}
procedure of {\em Pari-GP}.
\item In each of these cases, $L(x)$ is a monic polynomial with $L(0)=-1$
and hence $\lambda$ is a unit of the class field.
\item For $N=2317723$, the Hilbert class group
is cyclic and is generated by the equivalence class $[a,b,c]=[151,-91,3851]$,
with order $h(-N)=105$. In Section~5, I describe how 15,000 digits of
$\lambda$ were used to reduce it to a unit, which was then checked
at 40,000 digits of precision.
\item John Zucker and I have investigated some composite discriminants,
finding that $\lambda^2$ is a unit
when $N$ is a product of distinct primes greater than 3.
I have verified this for squarefree $N<2000$ with $N\equiv3$~mod~4
and coprime to 3. I have not yet found a simple criterion
that determines why $N=7\times11\times13\times19=19019$
yields $\lambda$ as a unit, while for $N=7\times11\times23=1771$
one must take $\lambda^2$ to form a unit.
\end{enumerate}

\section{Hilbert class field}

The Hilbert class field of $Q(\sqrt{-N})$ is generated by the
polynomial~\cite[Th.~7.2.14]{Cohen}
\begin{equation}
P(x)=\prod_{[a,b,c]\in H}\left(x-
j\left(\frac{b+\sqrt{-N}}{2a}\right)\right)
\label{Hilbert}
\end{equation}
where
\begin{equation}
j(z)=\left(\left(\frac{\eta(z/2)}{\eta(z)}\right)^{16}
+16\left(\frac{\eta(z)}{\eta(z/2)}\right)^8\right)^3.
\label{Klein}
\end{equation}
As shown in~\cite[Sect.~125, p.~461]{Weber},
a real root of $P(x)$ is supplied by
\begin{equation}
\left(\frac{256}{r^{16}}-r^8\right)^3=j\left(\frac{1+\sqrt{-N}}{2}\right).
\label{real}
\end{equation}
For $N=2317723$, $P(x)$ is a polynomial of degree 105,
whose integer coefficients have up to 3050 decimal digits,
making it rather difficult to reduce its roots to a set
of simple radicals.
Fortunately, we do not need to use $P(x)$. A more convenient
polynomial that generates the same number field will serve our purpose.
Using a class invariant defined in Section~4.2,
I found that the Hilbert class field
of $Q(\sqrt{-2317723})$ is generated by a compositum
of three polynomials that generate its sub-fields of prime degree, namely
\begin{eqnarray}
Q_7(x)&=&x^7-323x^5-6057x^4-35434x^3-186299x^2-1450032x-19143360\qquad{}
\label{Q7}\\
Q_5(y)&=&y^5-y^4-339y^3-7879y^2+146334y-566316
\label{Q5}\\
Q_3(z)&=&z^3-z^2-59z-322
\label{Q3}
\end{eqnarray}
where $Q_p$ has discriminant
\begin{equation}
D_p=f_p^2(-N)^{\frac{p-1}{2}}
\label{disc}
\end{equation}
with an index $f_p$. The indices
\begin{equation}
f_3=1,\qquad
f_5=2^{4}\times3\times5^2\times11\times17\times47,\qquad
f_7=2^{10}\times3^2\times19^2\times61\times f_5
\label{indices}
\end{equation}
fortunately contain no prime greater than 61.

\section{An efficient pair of class invariants}

The algebraic number $r$ in~(\ref{r}) is the real root of a
monic cubic polynomial
with coefficients in the Hilbert class field.
These coefficients are algebraically constrained by the
condition that~\cite{Weber}
\begin{equation}
\gamma_2=\frac{256}{r^{16}}-r^8
\label{gamma2}
\end{equation}
generates the Hilbert class field, while the minimal polynomial
for $r$ has degree $3h$ and generates a cubic relative extension.

For each of the 198 primes congruent to 3 modulo 8
and less than 6000, I found that
the cubic relative extension takes
the form~\cite{nmbrthry}
\begin{equation}
r^3-2(f r^2+g r+1)=0
\label{f-g}
\end{equation}
where $f$ and $g$ are algebraic integers of the Hilbert class field.
I then found that these algebraic integers obey the constraint
\begin{equation}
2f^4-16f^3g^2+20f^2g^4-12f^2g-8fg^6+16fg^3-2f+g^8-4g^5+3g^2=0
\label{zero}
\end{equation}
which indeed ensures that $r$ does not appear in
\begin{equation}
-\frac{\gamma_2}{32}=
8f^8+32f^6g+16f^5+40f^4g^2+32f^3g+16f^2g^3+6f^2+12fg^2+g^4+2g
\label{integer}
\end{equation}
as may be confirmed by using~(\ref{f-g}) to eliminate powers
$r^j$ with $j\ge3$ from~(\ref{gamma2}) and then using~(\ref{zero})
to eliminate powers $g^k$ with $k\ge8$.

A particularly simple example~\cite[Table VI, p.~725]{Weber}
is provided by $N=163$, the largest number for which $h(-N)=1$,
where the integer pair $[f,g]=[3,-2]$ determines the well-known
18-digit integer~\cite[Sect. 7.2.3]{Cohen}
\begin{equation}
-j\left(\frac{1+\sqrt{-163}}{2}\right)=262537412640768000
\label{j-163}
\end{equation}
that differs from $\exp(\pi\sqrt{163})$ by less than 3 parts
in $10^{30}$ and is here obtained by evaluating $-\gamma_2^3$,
using~(\ref{integer}). Hence $[f,g]=[3,-2]$ is a Diophantine
solution of~(\ref{zero}).

\subsection{A signature for $N\equiv3\mbox{ mod }8$}

I began my investigations by considering prime values of $N\equiv3$~mod~8,
since those yield a Chowla--Selberg unit, according to Conjecture~1.
Studying such primes, I discovered a signature, comprising a triplet of signs
$[S_1,S_2,S_3]$ that eventually enabled me to construct a pair of class
invariants for any number congruent to 3 modulo 8 and coprime to 3.

I arrived at this signature by using~(\ref{f-g}) to eliminate $f$
from~(\ref{zero}), obtaining an octic equation for $g$.
After some manipulations, I was able to solve this by taking 3 square roots.
The general solution for the octic has the form
\begin{equation}
g=-\frac{1}{r}
+S_1\left(r
+S_2\left(\frac{r^2}{2}
+S_3\left(\frac{r^{4}}{8}-\frac{1}{r^8}
\right)^{\frac12}
\right)^{\frac12}
\right)^{\frac12}
\label{signs}
\end{equation}
with signs $S_j=\pm1$.

By conjecture, precisely one of the 8 choices of signs
gives an algebraic integer of the Hilbert class field of $Q(\sqrt{-N})$.
If we know this signature, the problem of
identifying $k_N$ as an algebraic number becomes {\em much}
more tractable than previously supposed, since
instead of having to find an integer relation between
$3h+1$ numbers, namely $r$ and an integral basis
for a cubic relative extension of the Hilbert class field,
we now need a pair of relations between merely $h+2$ numbers,
namely $[f,g]$ and an integral basis for the Hilbert class field
itself. At large $N$, the coefficients in the minimal polynomial
of $g=O(\sqrt{r})$ have, typically, 48 times fewer digits than those
in the Hilbert polynomial~(\ref{Hilbert}).

I determined the signatures of the 198 primes
$N\equiv3$~mod~8 with $N<6000$ by trial and error,
using the {\tt lindep} procedure of {\em Pari-GP}
to search for a integer relation between the unique real
embedding of the integral basis {\tt nfinit(quadhilbert(-N)).zk}
and numerical evaluations of~(\ref{signs}) in each of 8 possible cases.
For each prime, I found precisely one valid signature.
Then I listed the first 12 primes for each signature, obtaining the sequences
\begin{eqnarray*}
[-1,-1,-1]&:&163,\,227,\,419,\,547,\,739,\,1123,\,1187,\,1571,\,
1699,\,2083,\,2339,\,2467\\{}
[-1,-1,+1]&:&11,\,139,\,331,\,523,\,587,\,907,\,971,\,1163,\,
1291,\,1483,\,1867,\,1931\\{}
[-1,+1,-1]&:&179,\,307,\,499,\,563,\,691,\,883,\,947,\,1459,\,
1523,\,1907,\,2099,\,2803\\{}
[-1,+1,+1]&:&59,\,251,\,379,\,443,\,571,\,827,\,1019,\,1531,\,
1723,\,1787,\,1979,\,2683\\{}
[+1,-1,-1]&:&3,\,67,\,131,\,643,\,1091,\,1283,\,1667,\,1987,\,
2179,\,2243,\,2371,\,2819\\{}
[+1,-1,+1]&:&43,\,107,\,491,\,619,\,683,\,811,\,1259,\,1451,\,
1579,\,2027,\,2347,\,2411\\{}
[+1,+1,-1]&:&19,\,83,\,211,\,467,\,659,\,787,\,1171,\,1427,\,
1619,\,1747,\,1811,\,2003\\{}
[+1,+1,+1]&:&283,\,347,\,859,\,1051,\,1307,\,1499,\,1627,\,
2011,\,2203,\,2267,\,2459,\,2843
\end{eqnarray*}
which led me to conjecture, as these 8 lists were slowly growing,
that the signature of a prime congruent to 3 modulo 8
is uniquely determined by its residue modulo 64,
as indeed turned out to be the case for the rest
of the sample of 198 primes.

I then checked that this is also
the case for all the composite integers less than 3500
that are congruent to 3 modulo 8 and coprime to 3,
using the {\tt nfisisom} routine of {\em Pari-GP}
in situations for which {\tt quadhilbert} did not furnish
a polynomial with a real root. (I thank Karim Belabas
for this workaround.)

Thus, for each square-free positive integer $N$
that is congruent to 3 modulo 8 and is coprime to 3
(and also for $N=3$ itself)
there appears to be a unique signature $[S_1,S_2,S_3]$,
determined by the residue of $N$ modulo 64,
such that~(\ref{signs}) yields an algebraic integer
of the class field.

\subsection{Construction and conjecture modulo 64}

For positive integer $N$ congruent to 3 modulo 8, I define
a signature
\begin{equation}
[S_1,S_2,S_3]=\left\{\begin{array}{l}
[-1,-1,-1]\mbox{ for }N\equiv35\mbox{ mod }64\\{}
[-1,-1,+1]\mbox{ for }N\equiv11\mbox{ mod }64\\{}
[-1,+1,-1]\mbox{ for }N\equiv51\mbox{ mod }64\\{}
[-1,+1,+1]\mbox{ for }N\equiv59\mbox{ mod }64\\{}
[+1,-1,-1]\mbox{ for }N\equiv 3\mbox{ mod }64\\{}
[+1,-1,+1]\mbox{ for }N\equiv43\mbox{ mod }64\\{}
[+1,+1,-1]\mbox{ for }N\equiv19\mbox{ mod }64\\{}
[+1,+1,+1]\mbox{ for }N\equiv27\mbox{ mod }64\end{array}\right.
\label{sig}
\end{equation}
and a pair of algebraic numbers
\begin{equation}
[f,\,g]=\left[\frac{r}{2}-\frac{s}{\sqrt{r}},\,
-\frac{1}{r}+s\sqrt{r}\right]
\label{pair}
\end{equation}
where
\begin{eqnarray}
r&=&\exp(-\pi{\rm i}/24)\,
\frac{\eta\left(\frac{1+\sqrt{-N}}{2}\right)}{\eta\left(\sqrt{-N}\right)}
\label{Weber-f}\\
s&=&S_1\left(1+S_2\left(\frac{1}{2}
+S_3\left(\frac{1}{8}-\frac{1}{r^{12}}
\right)^{\frac12}\right)^{\frac12}\right)^{\frac12}.
\label{s}
\end{eqnarray}
{\bf Conjecture 2}: For every square-free positive integer
$N$ congruent to 3 modulo 8 and coprime to 3,
the Hilbert class field of $Q(\sqrt{-N})$
is generated by at least one of $[f,g]$
and for $N>1099$ it is generated by both.\\[5pt]
Remarks:\begin{enumerate}
\item I have checked that the minimal polynomials of $f=f(N)$
and $g=g(N)$ have degree $h=h(-N)$ for all of the cases in Conjecture~2
with $1099<N<100000$.
\item There are 7 cases with $N\le1099$ in which only
one of $[f,g]$ generates the Hilbert class field, while the
other generates a sub-field.
\item Five of these yield the integers
$f(83)=1$, $f(91)=1$, $g(331)=-1$, $g(427)=1$, $g(907)=-2$
and were noted in~\cite{Russell}, with three cases appearing
in~\cite[Table~5]{Y-Z}.
\item For $N=715$, with $h=4$,
the minimal polynomial of $g$ is $x^2 + x - 1$.
\item For $N=1099$, with $h=6$, it is $x^3 + x^2 - x + 6$.
\item In the cases $N=11, 19, 43, 67, 163$, with $h=1$,
the $[f,g]$ pairs are $[1, -1]$, $[0, 1]$, $[1, 0]$, $[1, 1]$,
$[3, -2]$, all of which were noted in~\cite[Table~VI]{Weber}.
\item Apart from the 10 cases noted above,
no other value of $N<1000000$ produces an integer.
(The integers $f(3)=g(3)=f(27)=0$ do not fall within Conjecture~2.)
\item For $N<3500$, I have verified that whenever the minimal
polynomial of $f$ or $g$ has degree $h$ the field which it
generates is isomorphic to that generated by the
{\tt quadhilbert} procedure of {\em Pari-GP}.
\item I have performed the same tests for prime $N<6000$.
\item At large $N$, the minimal polynomial of $g$ provides a
rather economical generator of the field. For $N=2317723$,
it may be computed in less than 100 milliseconds and
has a height less than the {\em cube} root of the height of the
{\tt quadhilbert} polynomial.
\end{enumerate}

\subsection{Minimal polynomials}

The algebraic numbers $f$ and $g$ are, by construction,
roots of the polynomials
\begin{eqnarray}
F(x)&=&\prod_{j=1}^h
\left(x-\frac{r_{j,1}}{2}-\frac{r_{j,2}}{2}-\frac{r_{j,3}}{2}\right)
\label{F}\\
G(x)&=&\prod_{j=1}^h
\left(x+\frac{1}{r_{j,1}}+\frac{1}{r_{j,2}}+\frac{1}{r_{j,3}}\right)
\label{G}
\end{eqnarray}
where $r_{j,k}$ is a labelling of the roots
of the minimal polynomial of $r$ such that
\begin{equation}
\gamma_2\left(r_{j,1}\right)=
\gamma_2\left(r_{j,2}\right)=
\gamma_2\left(r_{j,3}\right)
\label{same}
\end{equation}
with $\gamma_2(r)=256/r^{16}-r^8$.
Conjecture~2 asserts, {\em inter alia},
that at least one of these polynomials
is irreducible and generates the Hilbert class field.

To compute the polynomials, we may use Reinier Br\"{o}ker's
fine formula~\cite[Th.~6.3, p.~106]{Broker} for the root
associated to the equivalence class $[a,b,c]$
of binary quadratic forms with discriminant $b^2-4ac=-4N$. Denoting
$z=(b/2+\sqrt{-N})/a$, this root is
\begin{equation}
R(a,b,c)=\left\{\begin{array}{ll}
-(-1)^{\frac{a^2-1}{8}}\exp\left(-\frac{b(a c^2-a-2c)}{48}\,\pi{\rm i}\right)
\,\frac{\eta(z/2)}{\eta(z)}&\mbox{ if $c$ is even}\\
-(-1)^{\frac{c^2-1}{8}}\exp\left(-\frac{b(c-a-5a c^2)}{48}\,\pi{\rm i}\right)
\,\sqrt{2}\,\frac{\eta(2z)}{\eta(z)}&\mbox{ if $a$ is even}\\
\exp\left(-\frac{b(c-a-a^2 c)+2}{48}\,\pi{\rm i}\right)
\,\frac{\eta((1+z)/2)}{\eta(z)}&\mbox{ otherwise}
\end{array}\right.
\label{R-a-b-c}
\end{equation}
where I have written the Weber functions as explicit eta quotients.
I remark that $R(1,0,N)=r$ determines the $N$th singular value~(\ref{r})
and that at least one of $[a,c]$ is odd, since $b$ is even.

When the class group for discriminant $-4N$ is generated
cyclicly, by a single class with order $3h$, there is a very simple
procedure to generate the roots with a labelling
that respects the condition~(\ref{same}): we may compute $r_{j,k}$
by applying~(\ref{R-a-b-c}) to the reduced form obtained
by raising the generator to the power $j+(k-1)h$. If there
are sub-groups, a little book-keeping is required to ensure that the
roots are slotted into~(\ref{F},\ref{G}) in a manner that respects
condition~(\ref{same}). I ordered the roots by size of the
real parts of their $\gamma_2$ values and then inspected the
signs of the imaginary part of $\gamma_2$

For $N>1099$, the minimal polynomial $G$ is a rather economical generator
of the Hilbert class field.
In the rather simple example of $N=1571$, with $h=17$, I obtained
\begin{eqnarray}
G(x)&=&x^{17} + 14x^{16} + 38x^{15} + 19x^{14} + 83x^{13}
+ 440x^{12} + 275x^{11} -507x^{10} + 384x^9\nonumber\\
&+& 541x^8 - 1343x^7 - 88x^6 + 712x^5 + 585x^4
- 1254x^3 + 852x^2 - 304x + 64\qquad{}
\label{min-g}
\end{eqnarray}
whose index
\begin{equation}
2^{37}\times13^2\times17^2\times41\times43\times139\times2083\times34259
=117388472496907896691997278208
\label{index-g}
\end{equation}
has merely 30 digits. By contrast the polynomial obtained
in~\cite[p.~152]{Broker}, using a double
eta-quotient~\cite{Schertz-units,Schertz,Enge-S} of the form
\begin{equation}
w_{p,q}(z)=\frac{\eta\left(\frac{z}{p}\right)\eta\left(\frac{z}{q}\right)}
{\eta(z)\eta\left(\frac{z}{p q}\right)},
\label{p-q}
\end{equation}
with $[p,q]=[5,7]$,
has a 52-digit index, while {\tt quadhilbert}
yields a 60-digit index, using $[p,q]=[29,31]$.

The economy of $G$ is also reflected
in the storage for the integral basis obtained by
outputting {\tt nfinit(G).zk} from {\em Pari-GP},
which produces a file of less than 12 kilobytes, while
{\tt nfinit(quadhilbert(-1571)).zk} produces more than 29 kilobytes.
This is because large divisors of the index occur in the
denominators of the rational elements of the matrix
that transforms powers of the root to an integral basis.

\section{Reduction to simple radicals for $N=2317723$}

For $N=2317723$, I used the generator $[a,b,c]=[604, 422, 3911]$,
with order 315, to obtain the polynomials $[F,G]$
from~(\ref{F},\ref{G}) in 90 milliseconds.
Their indices in the class field have
10,756 and 5,815 digits, respectively.
By way of comparison, the {\tt quadhilbert}
routine of {\em Pari-GP} gave an index with 20,075 digits.
The height of $G$ has 65 digits, while a 204-digit height was
produced by {\tt quadhilbert}. Using $G$,
I found the sub-fields~(\ref{Q7},\ref{Q5},\ref{Q3}).

\subsection{The elliptic integral $K_{2317723}$}

Inspired by the results in~\cite[pp.~238--247]{EMA},
obtained by Jon Borwein and John Zucker for
elliptic integrals $K_N$ with $N\le 100$,
my goal was to reduce the elliptic integral $K_{2317723}$
to $\Gamma$ values and the simplest possible radicals,
which I took to be those generated by the
polynomials $Q_7$, $Q_5$ and $Q_3$ in~(\ref{Q7},\ref{Q5},\ref{Q3}),
whose indices in sub-fields of the Hilbert class field contain no prime
greater than $61$. By contrast, a compositum of these polynomials
gave a 7419-digit index.

Nonetheless, I found it convenient to construct, for intermediate
purposes, a local integral basis from this compositum and then to use
{\tt lindep} to obtain the coefficients of $[f,g,\lambda]$
in this basis. The reason is simple: this is a triplet of
algebraic integers, so by using an integral basis we ensure that
no large denominator may leak into the $Q$-linear relations
and thereby inflate the typical size of numerators in
rational coefficients.

Hence the results were, in the first instance, in terms of a rather
unwieldy (yet computationally effective) integral basis,
occupying 74 Megabytes of disk space. However, it was possible
to shrink this data set, very dramatically.

\subsection{Reduction to monomials}

Next, I transformed $[f,g,\lambda]$ from the integral basis
to the 105 monomials $x^i y^j z^k$,
with $i<7$, $j<5$ and $k<3$, where $x$, $y$ and $z$
are the unique real roots of $Q_7(x)=0$, $Q_5(y)=0$ and $Q_3(z)=0$.
Then {\em Pari-GP} found that the {\tt content} of $g$ is $1/C$, where
\begin{equation}
C=2^{8}\times3^2\times5^3\times11^2\times17^2\times19^2
\times47^2\times61\times2317723=1135455149209896386784000
\label{C}
\end{equation}
has 25 digits. The resulting compact integer data for the
vector $V=[f,g,\lambda]$ is available (see the first footnote)
in the form of a 32-kilobyte file {\tt K2317723.txt}
that achieves a 2300-fold compression of the data from the integral basis.

I remark that my intermediate use of an integral basis had
the merit of reducing the working precision required for the reduction
of $\lambda$ to radicals by roughly 2,500 decimal digits,
i.e.~by about 25 digits per term in the reduction of the unit
$\lambda$ to an integral basis of the class field.

It seemed to me to be beyond reasonable expectation that {\em Pari-GP}
might determine a system of fundamental units
for the class field of $Q(\sqrt{-N})$ with $N=2317723$.
Hence I used only {\tt nfinit} at $N=2317723$,
while the more time-consuming procedure {\tt bnfinit} was used
to good effect for $N<6000$.

\subsection{Solution of sub-fields by radicals}

To complete the reduction to simple radicals,
I needed to determine the real roots
of the equations $Q_7(x)=0$, $Q_5(y)=0$, $Q_3(z)=0$
and then, from $f$ and $g$, the real root $r$ of the cubic~(\ref{f-g}).
It is elementary to solve a cubic by radicals. In particular,
\begin{equation}
z=\frac13
+\left(\frac{9227}{54}+\sqrt{\frac{2317723}{108}}\right)^{\frac13}
+\left(\frac{9227}{54}-\sqrt{\frac{2317723}{108}}\right)^{\frac13}
\label{z}
\end{equation}
is the unique real root of $Q_3(z)=0$.
To solve the quintic, we may compute the real parts
\begin{eqnarray}
u_n&=&\Re\bigg[
-(650272782-564880\sqrt{-2317723})\exp(2\pi{\rm i}n/5)\nonumber\\&&
-(1703074422-359490\sqrt{-2317723})\exp(4\pi{\rm i}n/5)\bigg]\label{u}
\end{eqnarray}
for $n=1\ldots4$, using $4\cos(\pi/5)=1+\sqrt{5}$. Then
\begin{equation}
y=\frac{1+u_1^{\frac15}+u_2^{\frac15}-(-u_3)^{\frac15}+u_4^{\frac15}}{5}
\label{y}
\end{equation}
is the unique real root of $Q_5(y)=0$.
To solve the septic, we may compute the real parts
\begin{eqnarray}
v_n&=&\Re\bigg[
-(1959346982341+140861987\sqrt{-2317723})\exp(2\pi{\rm i}n/7)\nonumber\\&&
-(686210881202-650234914\sqrt{-2317723})\exp(4\pi{\rm i}n/7)\nonumber\\&&
-(1670361863821+547274245\sqrt{-2317723})\exp(6\pi{\rm i}n/7)\bigg]\label{v}
\end{eqnarray}
for $n=1\ldots6$, using
\begin{equation}
6\cos(\pi/7)=1
+\left(\frac{-7+7\sqrt{-27}}{2}\right)^{\frac13}
+\left(\frac{-7-7\sqrt{-27}}{2}\right)^{\frac13}
\label{cos}
\end{equation}
and then
\begin{equation}
x=\frac{v_1^{\frac17}-(-v_2)^{\frac17}+v_3^{\frac17}
+v_4^{\frac17}+v_5^{\frac17}+v_6^{\frac17}}{7}
\label{x}
\end{equation}
is the unique real root of $Q_7(x)=0$.

The algebraic integers in $u_n$ and $v_n$
were found at 38-digit precision, using the method
outlined in~\cite[Chap.~3.1]{Morain} and there exemplified
by the quintic that generates the Hilbert class field
of $Q(\sqrt{-47})$.
As remarked in~\cite[VI-5]{CM}
that quintic
was solved by G.P.\ Young~\cite{Young} in 1888.
For G.N.\ Watson's comments on Young, see~\cite{B-S-W}.
For J.M.\ Whittaker's comments on Watson, see~\cite{Watson-obit}.
For the inspirational role of Srinivasa Ramanujan, see~\cite{B-C-Z}.

\subsection{Numerical checks}

At no stage in the reduction of $[f,g,\lambda]$ to such simple
radicals was it necessary to use a working precision
above 15,000 digits. The results were then checked at a precision
of 40,000 digits. For the singular value, that is very easy,
since we need only take seventh, fifth, cube and square roots
and check the relation between a pair AGMs in~(\ref{k-N}).
To check the elliptic integral, I evaluated the Chowla--Selberg
formula~(\ref{C-S}) at a precision of 40,000 digits.
As a final check that no stray
factor had been overlooked in going from the $\Gamma$ values
in~(\ref{G-N}) to the $\eta$ values in~(\ref{C-S}),
I evaluated 2,317,723 values of the $\Gamma$ function,
at 38-digit precision,
and combined them with the Kronecker symbol, obtaining
agreement with~(\ref{w}). The checking programme {\tt K2317723.gp}
and its output {\tt K2317723.out} are in the same directory as the
monomial coefficients, with a URL given in the first footnote.

\section{Comments and conclusion}

As announced in~\cite{nmbrthry,LL2008}, I had earlier reduced the
elliptic integrals $K_{34483}$ and $K_{1242763}$ to algebraic
numbers and $\Gamma$ values, following
the identification of elliptic integrals at singular values
in quantum field theory~\cite{B3G,contour}.
However, that was done more labouriously, without benefit of the
novel construction in~(\ref{sig}--\ref{s}).

The discoveries reported here stemmed from my persistent belief that
(notwithstanding well-intentioned advice to the contrary)
the problem of a polynomial with degree $3h$, for
singular values $k_N$ with $N\equiv3$~mod~8,
ought (at bottom) to be no more difficult than the problem with
degree $h$, for $N\equiv7$~mod~8.

It was thus rather gratifying to discover that 3~mod~8 is, in fact,
far preferable to 7~mod~8. In particular, I remark that:
\begin{enumerate}
\item The polynomial $G$ in~(\ref{G}) generates the Hilbert class field
with great (perhaps unprecedented) economy for large $N\equiv3$~mod~8
and coprime to 3, since it is precisely the trebling of roots
of the Weber polynomial that allowed me to combine their
reciprocals, three at a time. Thus we may avoid the large-$N$ growth
of $r=\exp(\pi\sqrt{N}/24)+o(1)$, using a level-48 class invariant with
growth
\begin{equation}
g=\alpha(N)\exp(\pi\sqrt{N}/48)+o(1)
\label{growth}
\end{equation}
where the asymptotic prefactor $\alpha(N)\in[-\sqrt2,\sqrt2]$ is given
by the signature~(\ref{sig}) as
\begin{equation}
\alpha(N)=\left\{\begin{array}{ll}{}
-\sqrt{1-\beta_-}\,=\,\sqrt2\cos(11\pi/16)&\mbox{ for }N\equiv35\mbox{ mod }64\\{}
-\sqrt{1-\beta_+}\,=\,\sqrt2\cos( 9\pi/16)&\mbox{ for }N\equiv11\mbox{ mod }64\\{}
-\sqrt{1+\beta_-}\,=\,\sqrt2\cos(13\pi/16)&\mbox{ for }N\equiv51\mbox{ mod }64\\{}
-\sqrt{1+\beta_+}\,=\,\sqrt2\cos(15\pi/16)&\mbox{ for }N\equiv59\mbox{ mod }64\\{}
+\sqrt{1-\beta_-}\,=\,\sqrt2\cos( 5\pi/16)&\mbox{ for }N\equiv 3\mbox{ mod }64\\{}
+\sqrt{1-\beta_+}\,=\,\sqrt2\cos( 7\pi/16)&\mbox{ for }N\equiv43\mbox{ mod }64\\{}
+\sqrt{1+\beta_-}\,=\,\sqrt2\cos( 3\pi/16)&\mbox{ for }N\equiv19\mbox{ mod }64\\{}
+\sqrt{1+\beta_+}\,=\,\sqrt2\cos(  \pi/16)&\mbox{ for }N\equiv27\mbox{ mod }64
\end{array}\right.
\label{alpha}
\end{equation}
with
\begin{equation}
\beta_{\pm}=\sqrt{\frac12\pm\sqrt{\frac18}}
\label{beta}
\end{equation}
obtained from~(\ref{s}) in the limit $r\to\infty$.
\item I find it notable that a novel solution to a problem relating
to elliptic integrals was suggested, almost by accident,
by typing merely 3 primes into Neil Sloane's wonderful search
engine for integer sequences~\cite{OEIS}, which shrewdly informed me
of a common residue.
\item The challenge of increasing the value $N$, of a square-free number
for which the complete elliptic integral $K_N$ has been successfully
reduced to explicit radicals and $\Gamma$ values, is now seen
to be {\em far} easier for $N\equiv3$~mod~8 than for $N\equiv7$~mod~8,
since the minimum value of $h(-N)$ accessible using the residue 3~mod~8
is approximately 3 times smaller than that for 7~mod~8,
for comparable $N$.
\item The cause is clear: we know the result for
the sum of Kronecker symbols in~\cite[Cor.~5.3.13]{Cohen}
\begin{equation}
\sum_{k=1}^{\frac{N-1}{2}}\left(\frac{-N}{k}\right)
=\left\{\begin{array}{rl}
3h(-N)&\mbox{for }N\equiv3\,{\rm mod}\,8\\
 h(-N)&\mbox{for }N\equiv7\,{\rm mod}\,8\end{array}\right.
\label{by-3}
\end{equation}
and have very little reason to expect the left-hand side
of this equation to favour one residue of $N$ over another, on average.
\item Indeed it does not. The smallest known odd class number $h(-N)$
for $N>2100000$ and $N\equiv3$~mod~8 is $h(-2317723)=105$, while
the smallest for $N\equiv7$~mod~8 is $h(-2140807)=309$. As
expected, from the right-hand side of~(\ref{by-3}), the latter
is close to 3 times former. It might have been thought, heretofore, that
what we gained on Kronecker's swings, by choosing 3~mod~8,
would be lost on Weber's
roundabouts so to speak\footnote{The colloquial saying seems to
be: ``What's lost upon the roundabouts, we pull up on the swings."},
where we are confronted by a Weber polynomial with degree $3h$
for the residue 3~mod~8.
\item However, I have demonstrated that nothing is lost, thanks
to the construction in~(\ref{sig}--\ref{s}) which
gives a pair of class invariants, both of whose minimal polynomials
have (conjecturally) degree $h$ for all square-free $N>1099$
with $N\equiv3$~mod~8 and $N$ coprime to 3.
One of these appears to outperform the double eta-quotient method.
\item It is understandable why the residue 3 modulo 8 was
discarded~\cite[Sect.\ 7.2.2, p.\ 46]{Atkin-M} in the early days
of elliptic curve primality proving: the factor $3$ in
the degree $3h$ of the Weber polynomial appeared to be a
considerable hindrance. Yet it is, in reality, a great {\em help} in
generating the class field of degree $h$, with true economy.
\item For $N=9760387\equiv3$~mod~8, mentioned in an
update~\cite[Table~3]{fastECPP} on progress~\cite{C-C}
with elliptic curve primality proving,
the minimal polynomial $G$ of the level-48 class invariant
$g$ in~(\ref{pair}) has a height whose {\em logarithm} is less than 37\%
of the logarithmic height generated by the double eta-quotient
used in {\em Pari-GP}. Moreover, the far simpler
polynomial $G$ was generated by~(\ref{R-a-b-c})
in less than 60\% of the time taken by
{\tt quadhilbert(-9760387)} in {\em Pari-GP}.
\item After completing this work, I found that the cubic
relative extension~(\ref{f-g}) had been analyzed
in~\cite{Russell,Watson-sing,Y-Z} in cases with class number $h(-N)\le5$.
\end{enumerate}

I conclude by remarking that
negative discriminants $D=-N$ with $N\equiv3$~mod~8
have recently been used to good effect in the
construction of elliptic curves of prime order~\cite{Broker}
as well as in elliptic curve primality proving~\cite{Primo,fastECPP}.
It may be that the class invariants $[f,g]$ constructed
in~(\ref{pair}) have something to offer researchers in these
and other fields. To that end, I append a polynomial, derived
from~(\ref{zero},\ref{integer}), that relates $g$ to the $j$-invariant.

\section{Appendix}
With $J=j((1+\sqrt{-N})/2)$ and $[f,g]$ defined in~(\ref{pair})
for $N\equiv3$~mod~8, I obtained
{\scriptsize\begin{eqnarray*}
&&4722366482869645213696g^{192}
+906694364710971881029632g^{189}\\&&{}
+83642555144587156024983552g^{186}
+4939066436035567493262082048g^{183}\\&&{}
+209846732144453295821190856704g^{180}
+6836790472875669456820597948416g^{177}\\&&{}
+177760660111365660798399713116160g^{174}
+3790405367998157254338394567213056g^{171}\\&&{}
+67599317184302478754990860798001152g^{168}
+1023330374861490173762756220786049024g^{165}\\&&{}
+13300167538995234485451503275754913792g^{162}
+149751357319880860617353032541637967872g^{159}\\&&{}
+1471242473645701356762195242184643444736g^{156}
+12686152623120457776559166922665911910400g^{153}\\&&{}
+96465713862314370555819332777575421313024g^{150}
+649387270593628934858171069925186898755584g^{147}\\&&{}
+(31230955333453581854030430208J
+3882453146659327990928087554832136180596736)g^{144}\\&&{}
+(3112575311968735739497345449984J
+20668528534099939341664223139973218586066944)g^{141}\\&&{}
+(146491081273850273193327964717056J
+98181174566531282177050821993847942140657664)g^{138}\\&&{}
+(4334567835473120225746709693595648J
+416862949310707523391004696461020551773683712)g^{135}\\&&{}
+(90573700669027853953409791435997184J
+1584081987345225328419300608733679840703545344)g^{132}\\&&{}
+(1423343973438783107899395237834915840J
+5392827180734138390120670122880709527544528896)g^{129}\\&&{}
+(17493275962926368294467182339498704896J
+16459794382816811643862933127629261242032455680)g^{126}\\&&{}
+(172648676792410562129703110693458280448J
+45061910572250059933411888109783903591347519488)g^{123}\\&&{}
+(1394283125785794590584373780949472641024J
+110684289672788641685181184738158837724230975488)g^{120}\\&&{}
+(9342192287286270079567239190370043559936J
+243939817239193661299082038559564687476650934272)g^{117}\\&&{}
+(52479612331578998117553933098199803756544J
+482336992597299364139466938834793244884021018624)g^{114}\\&&{}
+(249139788648436660109159830308175969517568J
+855388454309670556943328874006671088228310712320)g^{111}\\&&{}
+(1005715446292108629597040700247216865935360J
+1359908516549423968069669760282605503992983191552)g^{108}\\&&{}
+(3468496180439370615221376712976872425652224J
+1936722218974592826488979701653464696769575124992)g^{105}\\&&{}
+(10256774471149943627485447340354884299915264J
+2468005608190905568464422101020736836421097095168)g^{102}\\&&{}
+(26077249048140483395956375421712523558125568J
+2809141400052941240710393397868547027567342780416)g^{99}\\&&{}
+(74434605568023196281142352281600J^2
+57114891944394614356435459851412372728053760J\\&&{}
+2847596261406655579330927459837021685211463680000)g^{96}
+(2477554196157529183058923534417920J^2\\&&{}
+107908793049662694542903803183229293591265280J
+2557916576434113812829574035094728838099498958848)g^{93}\\&&{}
+(37777050131731872831172646094766080J^2
+176007812462339450094643683054165114857979904J\\&&{}
+2018094067316585479424095293307588682510015397888)g^{90}
+(350315392609385120206628736534052864J^2\\&&{}
+247919707199861830348868071022250230292676608J
+1375480968524779058279375724416284358713081331712)g^{87}\\&&{}
+(2212954931060628055518534721471512576J^2
+301530529184884050981024118577114960445308928J\\&&{}
+782912651925726045755575568707697403306226745344)g^{84}
+(10113655478296307547676853277516890112J^2\\&&{}
+316467192335142296745902365499603998948196352J
+342002381493088167709431759538867103984517120000)g^{81}\\&&{}
+(34686160560417913758497734880697253888J^2
+286296268001635388526891108596990972415442944J\\&&{}
+80670074215058098900212634195910829748908982272)g^{78}
+(91454819947811608348373674411331420160J^2\\&&{}
+222854932560540399953035451922185320440791040J
-32576013459072580759743422218374296690436866048)g^{75}\\&&{}
+(188538486053819166866726313869734576128J^2
+148855954137712460111900655737028200724692992J\\&&{}
-55144112538101344960539779845043749378943090688)g^{72}
+(307688231969241589959784881772377931776J^2\\&&{}
+84952975352400749448181433286283669083783168J
-39652008878894560091506319036249518347078598656)g^{69}\\&&{}
+(401243764881273752344676821704124661760J^2
+41126233908050171152515472191400945995743232J\\&&{}
-18737893972438524931145834580573490329621102592)g^{66}
+(421140463268410135952689102182263291904J^2\\&&{}
+16667390579331302318343971161533721635454976J
-5174077032622993644412698123126373574057656320)g^{63}\\&&{}
+(357737370032797306669416932583241940992J^2
+5506258143473539813062973804040786515329024J\\&&{}
+356706547547601649023766058702724252706013184)g^{60}
+(246929031550304841851998761113119358976J^2\\&&{}
+1391923871860051857163468765410511483305984J
+1324523691432456084761088145116209617060233216)g^{57}\\&&{}
+(138863115636906600095701410448505044992J^2
+218444669720290975310036847391841483489280J\\&&{}
+806529800268765371684542515696435618323103744)g^{54}
+(63697798479377155669175763462382419968J^2\\&&{}
-5462276197319471266819439416672115490816J
+266980239729130899108909660949569834722525184)g^{51}\\&&{}
+(698176579929963364344659968J^3
+23826118069257453400721993188748820480J^2\\&&{}
-13455662359513885691724184271536483467264J
+25376524169352783195973261633263764798177280)g^{48}\\&&{}
+(2550985327389109428468842496J^3
+7253560775325956917535952133876088832J^2\\&&{}
-1613591086910774630703152810299985231872J
-24065084574987736253843103599575646487969792)g^{45}\\&&{}
+(4174917207070705118814928896J^3
+1790785455303357581601357069218217984J^2\\&&{}
+1787431768056365804275779798027674320896J
-14211863708323917478342629778282407593508864)g^{42}\\&&{}
+(4040205466878976552201093120J^3
+356547255740949811729208872745828352J^2\\&&{}
+1125486365607802615981912345130657906688J
-3441206898378612596310745579245155244834816)g^{39}\\&&{}
+(2570697570474836080757047296J^3
+56806582737949670581168990750507008J^2\\&&{}
+226614072500397724462829613101904560128J
-112335117391643625302204517357601891024896)g^{36}\\&&{}
+(1131315371638572737196195840J^3
+7167453070918708350458449780277248J^2\\&&{}
-87239165909372962210788016406434676736J
+192249493128040752962179905607616239239168)g^{33}\\&&{}
+(352740455777859128457691136J^3
+706517742750046477475912431435776J^2\\&&{}
-64776713258997636457464555907753967616J
+66506990142748686364267453679736522276864)g^{30}\\&&{}
+(78521574805087093933473792J^3
+53633593765131740994404867899392J^2\\&&{}
-10038379384210031276177488830957355008J
+7410964829180529718510440126722403729408)g^{27}\\&&{}
+(12416913717118929289347072J^3
+3135707043050900857899932712960J^2\\&&{}
+1593542960922142737017862254701314048J
-659117895483763817020809120277963210752)g^{24}\\&&{}
+(1371359778179842251423744J^3
+108551055367292136549255217152J^2\\&&{}
+405810458730443210149413338518388736J
-482780631328626439347360470569198813184)g^{21}\\&&{}
+(102588738647821241548800J^3
+64988007965336090461895393280J^2\\&&{}
-1063100916132737610564329604644864J
-65330166994701834714296174655572017152)g^{18}\\&&{}
+(4951224026747224719360J^3
-38854100569579258684962242560J^2\\&&{}
-1959825089179216758355828729184256J
-2155988932398684231421595412269629440)g^{15}\\&&{}
+(142903607317254504448J^3
+7056016192482886441475506176J^2\\&&{}
-32318350469538093301391589113856J
+1183943345005433116201363571887570944)g^{12}\\&&{}
+(2182827387064418304J^3
-415803546176586840262311936J^2\\&&{}
+1949335548919313469500521709568J
+185598558328963647368433135255552000)g^9\\&&{}
+(14241167385034752J^3
+6745596914666936897372160J^2\\&&{}
-90486773832711112570699776000J
+9948227935453805037037289472000000)g^6\\&&{}
+(25348472307712J^3
-19845426622060560384000J^2\\&&{}
+4189061192520522792960000000J
+181543631801412552228864000000000)g^3\\&&{}
+J^4
+2654208000J^3
+2348273369088000000J^2
+692533995824480256000000000J\;=\;0\,.\end{eqnarray*}}

The corresponding polynomial relation for $f$ has degree 8
in $J$ and is likewise available in the file {\tt PhiJFG.txt}
with checks provided by {\tt PhiJFG.gp} and {\tt PhiJFG.out}
from the URL in the first footnote.

\section*{Acknowledgements}

I am very grateful to David Bailey, Karim Belabas,
John Bolton, Jon Borwein, Reinier Br\"{o}ker, Chris Caldwell,
Larry Glasser, Marcel Martin, Neil Sloane and John Zucker for their
generous advice and gentle encouragement.

\newpage
\raggedright
\end{document}